# DIVIDE ET IMPERA: MEMORYRANGER RUNS DRIVERS IN ISOLATED KERNEL SPACES

Igor Korkin, PhD
Security Researcher
Moscow, Russia
igor.korkin@gmail.com

**ABSTRACT**

One of the main issues in the OS security is to provide trusted code execution in an untrusted environment. During executing, kernel-mode drivers allocate and process memory data: OS internal structures, users' private information, and sensitive data of third-party drivers. All this data and the drivers code can be tampered with by kernel-mode malware. Microsoft security experts integrated new features to fill this gap, but they are not enough: allocated data can be stolen and patched and the driver's code can be dumped without any security reaction. The proposed hypervisor-based system (MemoryRanger) tackles this issue by executing drivers in separate kernel enclaves with specific memory attributes. MemoryRanger protects code and data using Intel VT-x and EPT features with low performance degradation on Windows 10 x64.

**Keywords**: hypervisor-based protection, Windows kernel, rootkits, attacks of memory, memory isolation.

## 1. INTRODUCTION

Microsoft Windows Operating System has dominated on the world's market of desktop and laptop computers for more than 30 years. Nowadays Windows OS is running on more than one billion computers in so many different fields: industries, banking, business and government, transport and logistics, research, you name it (Warren, 2018). Attacks on Windows OSes have always been a desirable goal for various malware and rootkits.

For more than 20 years Windows OS has run in a protected mode, provided by the architectures of the x86 and x64 processors. This mode includes several security features, one of those are four privileged levels to protect the system code and data from being overwritten by less privileged code. According to the Yosifovich, Ionescu, Russinovich, & Solomon (2017), Windows OS uses only two privileged levels of the protected mode: one for OS kernel and drivers (kernel mode) and the other one for user applications (user mode).

**Thread model and assumptions.** Currently, kernel-mode drivers share the same memory address space with the rest of the OS kernel. All drivers can read and write any part of kernel-mode memory without any hardware restrictions. This fact makes Windows OS to be prone to rootkit attacks and kernel exploitation, see Figure 1 (Oh, 2018).

These malware attacks leverage the same privileged level as the OS kernel and can be performed by the following (Desimone & Landau, 2018):

- installing signed malware drivers;
- exploiting driver vulnerabilities.

Using kernel-mode code facilities, intruders can achieve the following goals (Shirole, 2014):

- maintain hidden and long-term control of the infected computer,
- escalate privileges;
- steal users' data;
- disrupt industrial processes.

All of them can be achieved by illegal read and write access to the code of drivers' OS kernel as well as tampering with their allocated data. Recent examples of these attacks are here (Korkin, 2018-a).

In the last several years, Microsoft experts have integrated a number of security features to protect OS kernel from these attacks.

The oldest implementation is PatchGuard, which crashes the OS after revealing some changes of internal structures such as EPROCESS unlinking. A more recent one is Device Guard, which provides the integrity of all loaded drivers by clearing write attributes for all executable memory pages. Therefore, any changes of any code loaded into the kernel will crash the OS.

However, it is not enough to prevent kernel mode threats. Windows Security features provide neither integrity nor confidentiality of the allocated memory of the third-party drivers. In addition, the OS internal structures are not fully protected, and finally the drivers' code can be dumped and analyzed by reverse engineering.

Cyber security researchers are trying to fill this gap and propose various ideas.

The first memory isolation concept termed "Multics" have been proposed by Corbató & Vyssotsky (1965) more than half a century ago. It is based on creating memory regions with different access permissions and was implemented for the GE 645 mainframe computer.

Numerous security projects designed for modern Intel, AMD, and ARM CPUs have been presented and discussed for more than 10 years.

One recent example is AllMemPro system developed by Korkin (2018-a). This hypervisor-based system protects the data allocated by third-party drivers and prevents internal structures of Windows OS kernel from being patched. At the same time, AllMemPro does not provide code integrity and confidentiality.

This paper suggests a new vision of kernel-mode memory protection workflow. The idea is to execute each driver in a separate memory enclosure. Each enclosure has a specific memory configuration:

- a driver can access the memory pool, which it has allocated before;
- execution of all other drivers is blocked;
- read and write access to the memory allocated by other drivers is blocked.

While OS works the control has to be switched between these enclosures to provide:

- execution of drivers' code;
- prevention of illegal access to the code and data.

The idea of such isolated kernel spaces for drivers has been implemented in hypervisor-based system called MemoryRanger.

The remainder of the paper proceeds as follows:

Section 2 provides a detailed review of existing memory protection projects. It also contains the comparison table with about 30 research projects including Windows built-in solutions.

Section 3 contains the proposed way of applying Extended Page Tables (EPT) to isolate memory. This section includes the details of architecture and its implementation, benchmark results, and limitations outline of MemoryRanger.

Section 4 focuses on the main conclusions and further research directions.

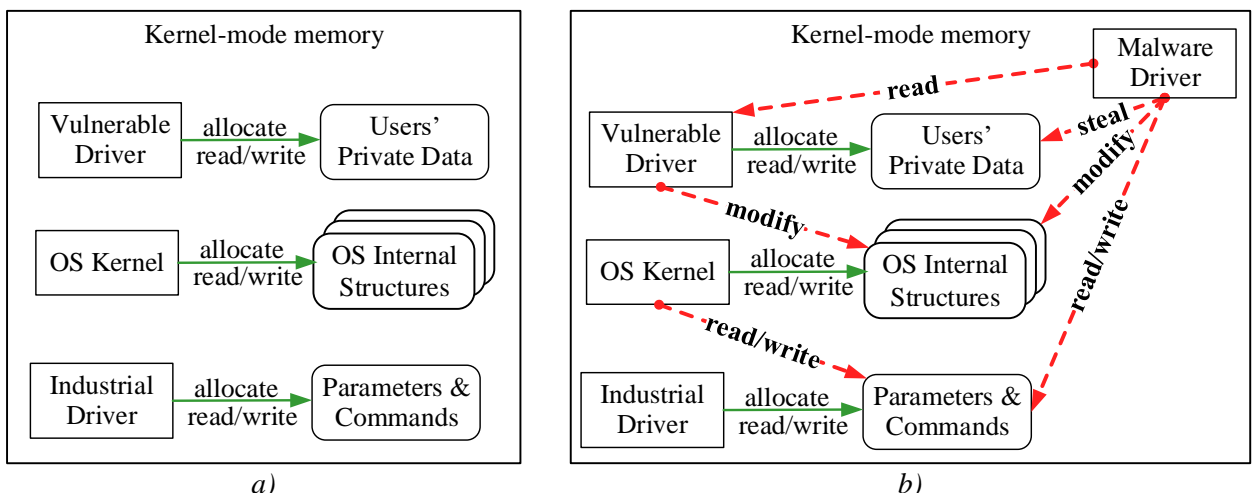

Figure 1 Examples of memory access attempts: *a)* without a malware driver and *b)* with a malware driver. Legitimate access attempts are illustrated as solid green arrows, unauthorized ones as dashed red arrows.

## 2. BACKGROUND

This section includes analysis of Windows built-in tools, research papers, and proof-of-concept prototypes that are designed to protect kernel-mode memory. All of them will be compared by their capabilities to provide both integrity and confidentiality for the code of OS and third-party drivers as well as for their dynamically allocated memory.

Since Windows XP, released more than 15 years ago and still is the 3rd most popular OS for PCs in the world (Ghosh, 2017), Microsoft has developed a stack of security components and integrated them into Windows 10. The components are the following: Smart Screen, Application Guard, Device Guard, Exploit Guard etc. (ACSC, 2018).

The Kernel Mode Code Integrity (KMCI), which is a part of Device Guard, prevents code modification of OS kernel and third-party drivers. It automatically resets the write permission bits for executable memory pages, so any code modifications will immediately cause BSOD with bug check code 0xBE.

Another component is PatchGuard, which protects the data of internal structures. It checks the integrity of some fields of these structures and causes BSOD with bug check code 0x109 after revealing an illegal modification. PatchGuard prevents, for example, EPROCESS unlinking, but skips privileges escalation attacks (Korkin, 2018-a).

The detailed analysis of Device Guard and PatchGuard is here (Korkin, 2018-a). Windows security features do not protect memory allocated by third-party drivers from being read or overwritten, and only partially provide integrity of internal Windows structures, see Table 1.

Security researchers from all around the world are trying to fill this gap and protect kernel-mode memory. An analysis of projects and prototypes will be given according to which memory class and from which type of access they protect:

- a code or assemblies of OS and third-party drivers from illegal read- and write- access;
- data allocated by OS and internal structures lists from illegal read- and write- access;
- data allocated by third-party drivers from illegal read- and write- access.

There are a lot of methods to monitor and prevent memory access: hardware-based and software ones; based on OS internal features and based on bare-metal hypervisors. The detailed analysis of all these methods are presented in Section 2 Korkin & Tanda, (2017).

This paper is primarily focused on the hypervisor-based methods, because they have several competitive advantages: they are fast, resilient to kernel-mode attacks; and work on all modern computers.

Yi et.al. (2017) proposed a fast data anomaly detection engine for kernel integrity monitoring called DADE. One of the authors' motivation is to prevent the adversary from obtaining the highest privilege by attacking the kernel. The main idea is to trap memory modifications and check their eligibility using backtraces. These backtraces include kernel functions that are used to create kernel-mode objects. DADE sets the whole kernel memory as read only so any write toward a kernel memory page generates a page fault, which is handled by the hypervisor. To achieve it DADE hypervisor leverages Extended Page Tables (EPT) functionality supported by hardware virtualization extensions. DADE assumes that of OS kernel source code is available and it supports only Linux OS. DADE prototype was integrated to a KVM with Linux OS and tested on ARM Cortex-A15.

Another hypervisor-based system has been proposed by Wang et.al. (2017). Their hypervisor-based access control strategy (HACS) protects security-critical kernel data of Linux OS kernel. This hypervisor plays the role of PatchGuard but for Ubuntu OS without crashing the OS after revealing data modification. HACS detects and prevents rootkits with DKOM and Hijack system calls attacks. It protects both types of data: static kernel data and dynamically allocated kernel data using module with whitelist-based access control. The key feature of HACS is that it can detect when rootkits place malicious code in module's initialization function (.init.text section). HACS leverages EPT mechanism to protect memory. It prohibits illegal modifications of sensitive data by marking the corresponding memory pages as read-only. HACS hypervisor traps memory access violations that occur during write access on these pages. After that to process write request, it takes advantage of the processor's single-step interruption mechanism. For the legal request this single step helps to allow write access just for one instruction. HACS uses only one EPT structure to protect OS kernel data. This system needs to intercept both legal and illegal access which is time consuming. It is implemented on BitVisor for the protection of the Ubuntu OS running on an Intel Core i7-4790.

Manes et.al. (2018) addresses the problem of rootkit attacks and kernel exploitation by presenting a new kernel architecture which securely isolates the untrusted kernel extensions. The implemented Domain Isolated Kernel (DIKernel) prevents three commonly used rootkit techniques:

- inline hooking;
- function pointer hooking;
- DKOM attacks.

DIKernel enforces memory access isolation by separating the kernel extensions from the rest of the OS kernel. DIKernel leverages the Domain Access Control Register (DACR), which is an ARM CPU hardware feature. The authors underline that using this feature makes it possible to organize memory domains with a very quick switch between them. DIKernel is implemented on Linux OS and tested on a Raspberry Pi 2 model B.

Another research focuses on preventing privilege escalation attacks (Qiang et.al., 2018). These attacks manipulate security-sensitive data in the kernel by exploiting memory corruption vulnerabilities. The authors underline that these non-control-data attacks are able to bypass current defense mechanisms. The proposed system called PrivGuard monitors the change of the sensitive kernel data by modifying the system call entry point, before system call entering, and before system call returning. PrivGuard is only based on OS internal mechanisms without using any hypervisor facilities. PrivGuard prototype is implemented on Ubuntu OS using x86 architecture.

Brookes et.al. (2016) consider attacks which involve dynamically disassembling kernel code stored in memory, privilege escalation, exploit artifacts management, and hiding behavior. To mitigate these memory disclosure attacks, the authors developed ExOShim, a hypervisor-based system, which renders all kernel code execute-only. The system leverages VT-x and EPT so that the hypervisor can install itself under a running kernel, which makes it possible to provide the complete execute-only memory access control primitives on all kernel code pages. By using EPT feature, ExOShim loads kernel code on memory frames that are marked as non-readable, non-writeable, but also executable. This is the entire premise of ExOShim. Another feature of ExOShim is self-protection, which prevents its code and data from any read, write, and execute access. As a result, an attacker cannot overcome the protection mechanism even trying to install malware hypervisor. ExOShim is a lightweight hypervisor for Windows-based OS, tested on Intel Core i7-3770k. ExOShim prevents kernel-mode code from illegal read and write access, and it protects only data needed to maintain ExOShim. It does not protect kernel-mode data of OS and third-party drivers.

The issue of providing integrity for the OS kernel data structures is considered by the security experts from the HP Labs and University of Texas at Austin (Hofmann et.al., 2011). The authors proposed a hypervisor-based system OSsk to prevent several rootkits attacks: hiding a rootkit process by removing its structure from the list and changing function pointers to the custom functions. The developed prototype OSsk protects the kernel data structures by verifying their content in a thread that runs concurrently with the guest execution. OSck protects the system call table through hardware page protection and a hypervisor call ensuring that once the table is initialized, the guest OS may not modify it. OSsk is implemented as a part of KVM hypervisor on the Intel Core i7 860 using Linux OS.

Another project leveraged hardware-virtualization technology and EPT feature of CPU is the InkTag hypervisor (Hofmann et.al., 2013). It allows executing trusted user-mode applications under an untrusted operating system. InkTag runs trusted applications in a special high-assurance process (HAP) which is isolated from the OS. InkTag provides a special hypercall to HAP to verify the runtime behavior of the OS. InkTag hypervisor ensures privacy and integrity for the code and data in a HAP's address space through encryption and hashing, and verifies that those services work correctly. InkTag is developed on the top of KVM hypervisor and tested using Linux OS on an Intel

Core i7 860. Although InkTag does not prevent any attacks on kernel memory, its concept seems very interesting and promising for OS security.

Security researchers from China proposed a pattern-based system to check integrity of the OS kernel (Feng et.al., 2018). Their system BehaviorKI extracts a set of patterns which characterize malicious behaviors. During the implementation of BehaviorKI the authors utilize hardware-assisted virtualization and EPT features designed for memory virtualization. Their behavior-triggered system inspects whether the OS critical components are modified illegally including static and dynamic components. The authors designed a testbed to imitate malicious behavior by extracting frequent event sequences from malware attacks. The researchers underline that "dynamic non-control data structures store critical information and user identification data" and emphasize that "integrity is very important to the security of the computer system". To reduce the performance penalty, BehaviorKI controls memory access events only for critical memory regions. This system also controls register operations and those involving system calls. BehaviorKI provides integrity for the following OS components: kernel code and static kernel data, dynamic kernel data including control-flow and non-control data. One example of dynamic kernel data is the head of LKM list, which is used by rootkits to hide themselves. To verify whether kernel integrity has been tampered, BehaviorKI uses kernel data invariants defined via kernel source code analysis and OS runtime snapshots. This system intercepts memory access operations by utilizing EPT violations and "removing the readable or writable permission to the monitored memory pages from the EPT entry". BehaviorKI processes these events by setting Trap Flag (TF) and recovers these pages to be readable and writable. During dispatching, the trap debug exception BehaviorKI blocks the permission to the monitored pages and clears TF to make the system run normally. The key feature of BehaviorKI is that it triggers integrity checking only when the event sequences match malicious behavior patterns. As a result, it has a lower performance overhead due to integrity checking compared to event-triggered approaches. The authors implemented BehaviorKI prototype on top of Xen hypervisor and tested it using the Intel Core i7-4710MQ and Linux OS.

LKMG is one of the most long-running research projects, dealing with the OS kernel protection against vulnerable loadable kernel modules (LMG). The researchers from the USA and China have been developing their prototype of LKM guard (LKMG) for more than 7 years (Tian et.al., 2011; Tian et.al., 2018). The authors consider that vulnerable LKM can modify any kernel data and code, call kernel functions, and read sensitive information. To protect OS from these attacks the authors utilize static analysis to extract the kernel code and data access patterns from kernel module's source codes and then generate a security policy by combining these patterns with the related memory address information. Also the authors isolate kernel module from the rest of the OS and enforce its execution by using hardware virtualization technology. The security policy is developed according to the principle of least privilege: an LKM can only access the kernel data that are necessary for its functioning. LKMG applies two EPT structures to mediate memory access: one EPT for the LKM and another one for the OS kernel. LKMG supports allocated memory protection by intercepting the allocation and deallocation functions. For the dynamic allocated data, LKMG prevents illegal read- and write- access. Another LKMG feature is kernel stack protection. LKMG protects the data allocated by the OS from being read and overwritten by LKM. It also guarantees the integrity of the OS kernel code. However, LKMG does not restrict the OS kernel, it can read and modify LKM code and data. One of the main issues of this guard is that it requires OS and LKM source code and does not support kernel protection when its source code is not available. The proposed policy-centric approach is implemented using Xen hypervisor and is developed for Linux OS running on Intel Xeon X3430. The authors assume that LKMG can be applied for Windows-based OS also.

Another project that performs integrity checking of the Linux OS kernel and has a similar name with the previous one is Linux Kernel Runtime Guard, or LKRG, developed by Zabrocki (2018). This project is designed to protect OS kernel against attacks via kernel vulnerabilities. LKRG performs post-detection and responds to unauthorized modifications of the Linux OS kernel and processes credentials. This guard provides OS kernel integrity and exploit detection. LKRG is currently in an early experimental stage of the development and it seems quite promising.

An interesting idea of providing integrity of the OS kernel code and data suggested by Kwon et.al. (2018). To avoid two-stage paging overhead authors implemented Hypernel security framework which relies on special hardware as well as a new software module called Hypersec. The hardware module called memory bus monitor (MBM), connects to the system bus between the CPU and main memory. MBM monitors write operations to every memory word and raises an interrupt upon finding any write attempt to the sensitive data. However, this bus monitor cannot be aware of the memory addresses of dynamically allocated kernel data objects, it can only be applied for monitoring limited kernel objects. The Hypernel prototype is implemented on the Versatile Express Juno r1 Platform running Linux OS.

EPTI developed by Hua et.el. (2018) is one of the recent hypervisor-based projects focused on kernel-mode memory leakage prevention. This project deals with the protection of the cloud computing systems against Meltdown Attack. This recently discovered attack makes it possible to dump the kernel code and data from user-mode applications. One of the main features of EPTI is the allocatation and switching between two EPT structures to isolate user space and kernel space. Another key feature is that EPTI overhead is quite low because each EPT structure has its own TLB and as a result, switching between EPTs does not flush TLB. The authors reveal one interesting fact during performing a real Meltdown Attack: they found "that although Meltdown can read the memory without access permission it cannot fetch code without executable permission even in reorder-execution". EPTI leverages this fact in the following way: all user memory has been mapped as execute-never in the corresponding EPT. EPTI prototype has been implemented on Linux OS running on Intel Core i7-7700.

Providing kernel integrity is also important even for smartphone OSes running on ARM CPUs. Ge, et.al. (2014) assume that OS kernel includes at least one exploitable vulnerability, which can be used by an adversary to hijack the control flow or to launch ROP attack. To prevent this attack the Sprobes system has been designed. This system provides kernel mode code integrity and prevents memory modifications caused by rootkits by using hardware extension ARM TrustZone. This makes it possible to partition all system resources and protect the confidentiality and integrity of all computations. Sprobes is implemented for Linux OS running on ARM Cortex M15.

The idea of using several different EPT paging structures to protect critical memory areas from kernel-mode malware has been implemented in the LAKEED system by Tian et.al. (2017). This system is specifically designed to prevent the kernel-level keyloggers from accessing the user buffer that contains the keystrokes. Authors assume that Windows OS kernel mainly utilizes two kernel modules to drive a keyboard, and to protect the page with keystrokes, they allocate three separate EPTs for the two keyboard drivers and one for the target kernel extension. All these three EPTs have the same memory mapping but with different access permissions: the target extension can only access its own memory region and cannot access the code and data region of the keyboard drivers. In the keyboard driver spaces both of the two drivers can access each other's data region in addition to their own code and data regions. LAKEED also prevents drivers' code mutual access as well as preventing access to the code and data of the target kernel extension. LAKEED is tested using Intel Xeon E5606 CPU for Windows OS. LAKEED protects limited kernel data buffers, which are related only to the keystrokes but demonstrates the possibility of using hypervisor with EPT support to isolate both code and data. One of interesting facts of LAKEED implementation is that it works well with filter drivers with minimal overhead. This fact demonstrates the possibility to use various EPT structures to isolate all filter drivers.

He et.al. (2017) is concerned about the security issue of attacking sensitive applications by exploiting kernel vulnerabilities. This issue makes sense for the current OSes, which use large monolithic kernel, because the kernel has complete access and control to/over all system resources including memory, device, and file management. In order to prevent kernel-mode attacks, the authors proposed a security-sensitive application (SSApp) protection mechanism called TZ-SSAP. This secure mechanism is based on a hardware-assisted environment provided by TrustZone technology on ARM CPU. TZ-SSAP protects the code integrity by modifying the access permissions for kernel memory pages and traps all updates. TZ-SSAP protects the integrity and security of kernel data: it maps static data with read-only attributes as the kernel mode; and makes dynamic data write-protected since it may be changed. TZ-SSAP has been tested on malware LKM that tries to directly tamper code and static data in the kernel space, as well as taking advantage of its privileges. The experimental results show that TZ-SSAP can prevent those attacks quite effectively. This system has been implemented for the ARM-Linux OS running on the ARM CoreTile Express A9x4 board.

A recently presented research project at the Black Hat USA 2017 conference focused on preventing modern OSes from being exploited using ROP payload, "just-in-time" ROP or JIT-ROP (Pomonis et.al., 2017). During these attacka, the exploit pinpoints the exact location of ROP gadget and assembles them on-the-fly into a functional JIT-ROP. The authors assume unprivileged local attacks, which may overwrite kernel code pointers with the OS via buggy kernel interfaces. To prevent these attacks, they present a kernel hardening scheme based on execute-only memory and code diversification called kR^X. This system includes two main parts: the R^X policy, and fine-grained KASLR. The R^X memory policy imposes the following property: memory can be either readable or executable. Fine-grained KASLR refers to a set of code diversification techniques specifically tailored to the kernel setting. This system helps to provide self-protection of execute-only kernel memory. kR^X prototype is implemented for the x86-64 Linux OS running on Intel Core i7-6700K CPU.

The issues of detecting illegal memory access have been considered in the system DigTool which is designed to detect various kernel-mode vulnerabilities (Pan et.al., 2017). This system can identify out-of-bounds, use-after-free, and time-of-check-to-time-of-use vulnerabilities for both kernel code and device drivers for Windows 7 and 10. DigTool leverages hypervisor facilities to monitor memory access by clearing the present flag (P flag) on the pages, which need to be monitored and processes page fault exceptions, which are triggered after any access to this page. DigTool enables the "single-step" operation by setting MTF (or TF) to trace access to this page. The authors focus on the two goals of illegal memory abuse: accessing beyond the bounds of the allocated heaps and referencing to already freed memory. To process all these events, DigTool hooks Windows allocation and deallocation functions: ExAllocatePoolWithTag, ExFreePoolWithTag, RtlAllocateHeap, and RtlFreeHeap. To prevent specifically UAF vulnerability DigTool also hooks InterlockedPushEntrySList and InterlockedPopEntrySList to be able to monitor the freed memory blocks in the lookaside lists. DigTool prevents read and write access to the freed memory and bounds of allocated pools.

Wang et.al. (2015) highlighted that modern OS kernels are too complicated to be secure because they consist of tens of millions of lines of source code. Consequently, an increasingly large number of vulnerabilities are discovered in all major OS kernels each year. The proposed SecPod framework provides a trusted execution environment by creating a dedicated address space or secure space in parallel to the existing kernel address space or the normal space. The authors considered attacks resulting in illegal modifications of the page table attributes. The secure space is designed to enforce memory isolation by sanitizing the guest page table updates. The authors have tested their solution using two various attacks: under execution of unauthorized code and under tampering data. To deal with the first scenario SecPod registers a call back function for kernel page updates. As a result, after a new executable page is created in the kernel mode, SecPod verifies the hashes of code pages. If the verification is correct, the page is marked as executable in the shadow page table. Otherwise, it detects an attempt to execute an unauthorized kernel code. For the second scenario, SecPod applies data invariants, which are used to prevent kernel data structures from being intricately interconnected. SecPod is developed using KVM hypervisor and the EPT feature, which the authors called NPT. This system is tested using Ubuntu OS running on Intel Core i5 CPU.

Hypervisor-based security solutions can be used to protect against physical attacks on main memory, such as cold boot attacks (Götzfried et.al., 2016). Intel SGX is one of the newly integrated technologies focused on the restriction of memory access. This technology leverages enclaves, but it is only suitable to protect user mode memory. Kernel mode space cannot be protected by Intel's SGX. To tackle this issue the authors developed HyperCrypt, which prevents physical attacks on kernel and user memory by leveraging VT-x with EPT (SLAT) technologies and AES symmetric encryption. HyperCrypt encrypts host physical pages and automatically decrypts pages that are currently accessed by the guest OS. HyperCrypt uses the CPU-bound encryption principle to prevent cryptographic keys and key material from being stored in RAM. HyperCrypt prototype is developed on top of BitVisor and is tested on Linux OS. HyperCrypt is not designed to prevent kernel memory from the kernel code access. Still the idea to prevent physical attacks, such as cold boot, using only software facilities seems very promising.

Srivastava et.al. (2009) underline that kernel components of the commodity monolithic OS, like Windows and Linux, share a unified address space that allows any component to access the data and code. Malicious kernel-level components can hide their own presence by illicitly removing OS data structures and can escalate process privileges by overwriting the process' user credentials with those for a root or administrative user. The authors presented the system called Sentry, which prevents kernel components with low trust by altering security critical data used by the kernel. Their system focuses primarily on the protection of dynamically allocated data structures. The authors assume that the core kernel code is protected and cannot be subverted by any malware. Their hypervisor-based design mediates the memory access attempts to overwrite protected data into the kernel address space. Sentry partitions kernel memory into two parts: protected and unprotected. It mediates memory access by using memory page protection bits. As a result, both legitimate and malicious writes to the protected pages will cause a page fault received by Sentry. Sentry determines the initiator of the access to protected data by using records on the kernel stack at every access. Its mechanism verifies memory access at the granularity of high-level language variables in the kernel's source code. Sentry does not provide privacy of allocated data and of the drivers' code and requires the drivers' source code. Sentry is developed using XEN and Linux OS kernel.

Lin et.al. (2007) underline that the primary cause of most OS failures are errors in device drivers written by third-party vendors. They point out that a malicious device driver can crash the whole OS or compromise its integrity because of unrestricted access to its resources. They proposed a system called iKernel to protect a kernel OS from both buggy and malicious device drivers. This system is designed to provide strong isolation mechanisms for device drivers using hypervisor-assisted virtual machine technology. The idea behind is the device drives isolation using separate virtual machines. As a result, the crash of the driver will affect only its virtual machine without affecting the host kernel or other virtual machines. This system isolates access to I/O ports and memory-mapped registers. iKernel does not provide integrity and privacy for the data allocated by the third-party driver. The system is developed using KVM virtual machine and Linux OS.

The research work presented by Chen et.al. (2017) is focused on the fact that commodity OS kernels are typically implemented using low-level unsafe languages. As a result, memory corruption vulnerabilities are quite common and inevitable security weakness of modern OS kernels. Their research considers memory corruption of non-control kernel data such as process credentials data. The system PrivWatcher is developed to protect the integrity of process credentials from these attacks using dual reference monitor which guarantees the Time-of-Check-To-Time-of-Use (TOCTTOU) consistency protected data fields. This system provides Discretionary Access Control (DAC) policy and prevents unauthorized processes from elevating their privileges. PrivWatcher does not protect kernel code and data allocated by third-party drivers from being read or patched. This framework has been implemented using Linux OS kernel with KSM.

Another security system presented by Azab et.el. (2014) relies on hardware features of the ARM to prevent attacks that aim to do the following:
- inject malicious code into the kernel;
- modify privileged code binaries that already exist in memory.

The proposed TrustZone-based Real-time Kernel Protection (TZ-RKP) provides OS kernel isolation using ARM Trust-Zone. TZ-RKP completely protects the kernel code base, but does not prevent attacks that trick the kernel into maliciously modifying its own data. This system is implemented using Android's Linux Kernel.

One more security system focused on providing code integrity of software running on commodity hardware has been presented by Zhang et.al. (2014). The proposed HyperCheck is a hardware-assisted integrity monitor, which successfully detects rootkits and code integrity attacks. HyperCheck prevents attacks on both targets: OS kernels, such as Linux and Windows and hypervisors, such as XEN. HyperCheck relies on the CPU System Management Mode (SMM) to acquire and transmit the state of the protected machine to an external machine. This system guarantees OS kernel code integrity only, without providing any protection against allocated data.

A group of security experts from Belgium (Gadaleta et.al., 2012) presented HyperForce, a hypervisor-based framework, which guarantees the execution of critical code in the kernel-space regardless of the state of the kernel, even if the OS kernel has been compromised. The authors assumed that a kernel-level attacker can modify the critical code in order to compromise its efficiency or completely disable its operations. To provide kernel-mode code integrity, HyperForce takes the advantages of hardware-based virtualization and write-protects the memory pages holding the instructions and data of the security-critical code. This framework allows the code to make changes to its data by unlocking the memory pages before it triggers an interrupt, and then lock them back immediately after the code's execution. HyperForce also write-protects the memory holding the Interrupt Descriptor Table (IDT) and protects the Interrupt Descriptor Table Register (IDTR) that contains the address of the IDT. HyperForce has been implemented using KVM and Linux OS kernel. HyperForce does not protect third-party drivers and memory allocated by them.

Another hypervisor-based framework was presented at Black Hat Asia by Han et.al. (2017). The proposed integrity protector Shadow-box supports periodic and event-based monitoring of kernel objects. Shadow-box recognizes integrity breaches in static and dynamic kernel objects. Shadow-box relies upon its two sub-parts: a lightweight hypervisor (Light-box) and a security monitor (Shadow-watcher). The Light-box is a lightweight hypervisor, which isolates OS kernel and dynamic kernel objects. The security monitor Shadow-watcher monitors static kernel elements and checks the integrity of dynamic kernel elements. Shadow-box protects the integrity of static kernel objects: code and data by setting *read and execute* rights for the code and only *read* rights for read-only data. This framework does not provide code privacy as well as security for the third-party drivers code and their data.

To protect commodity OS kernels from untrusted kernel extensions Xi, Tian, & Liu (2011) proposed HUKO, a hypervisor-based integrity protection system. This system leverages mandatory access control policies to limit an attacker's ability to compromise the kernel integrity. HUKO protects code, static data and dynamic data of the OS kernel from being modified.

SIDE is another system, which isolates the kernel from buggy device drivers developed by Sun and Chiueh (2011). SIDE executes a device driver in the same address space as the kernel but in a different protection domain from the kernel.

**Conclusion.** The conducted analysis of the related research projects shows that there are numerous hardware-based virtualization prototypes for Linux and Windows OS designed to prevent malicious kernel mode code from accessing code and sensitive data in the kernel memory. At the same time, not one of the existing solutions provide code and data protection for both OS kernel and third-party drivers, see Table 1.

Several projects leverage Intel EPT technology to create isolated enclaves for drivers, for example:

- one EPT used in HACS by Wang et.al. (2017), and AllMemPro by Korkin (2018-a);
- two EPTs used in LKMG by Tian et.al. (2018), and EPTI by Hua et.el. (2018);
- three EPTs used in LAKEED by Tian et.al. (2017);

EPT technology seems very capable of creating isolated kernel spaces; the detailed analysis of this possibility will be presented further.

Table 1. Summary table of memory protection projects

| Title, year | Unauthorized access to the following memory: | | | | | | OS |
| --- | --- | --- | --- | --- | --- | --- | --- |
| | Code for OS Kernel and 3rd party drivers | | Data of OS internal structures | | Data of 3rd party drivers | | |
| | Read | Write | Read | Write | Read | Write | |
| Windows built-in Protection, 2018[1] | – | + | – | + | – | – | Windows |
| Multics, 1965[2] | – | + | – | + | – | – | GE 645 |
| Sentry, 2009 | – | + | – | + | – | + | Linux |
| iKernel, 2007 | – | + | – | + | – | – | Linux |
| HUKO, 2011 | – | + | – | + | – | – | Linux, Windows |
| OSsk, 2011 | – | + | – | + | – | – | Linux |
| HyperForce, 2012 | – | + | – | – | – | – | Linux |
| HyperCheck, 2014 | – | + | – | – | – | – | Linux, Windows |
| HyperCrypt, 2016[3] | – | – | – | – | – | – | Linux |
| Hypernel, 2018[4] | – | + | – | + | – | – | Linux |
| PrivWatcher, 2017 | – | – | – | + | – | – | Linux |
| PrivGuard, 2018[5] | – | – | – | + | – | – | Linux |
| TZ-RKP, 2014 | – | + | – | – | – | – | Linux |
| TZ-SSAP, 2017 | + | – | – | – | – | – | Linux |
| SIDE, 2013 | – | + | – | + | – | + | Linux |
| Sprobes, 2014 | – | + | – | – | – | – | Linux |
| SecPod, 2015 | – | – | – | + | – | – | Linux |
| ExOShim, 2016 | + | + | – | – | – | – | Windows |
| LAKEED, 2016[6] | + | + | + | + | – | – | Windows |
| Shadow-box, 2017 | – | + | – | + | – | – | Linux |
| kR^X, 2017 | + | + | – | – | – | – | Linux |
| Digtool, 2017[7] | – | – | + | + | – | – | Windows |
| DADE, 2017 | – | + | – | + | – | – | Linux |
| HACS, 2017 | – | – | – | + | – | – | Linux |
| DIKernel, 2018 | – | + | – | + | – | – | Linux |
| BehaviourKI, 2018 | – | + | – | + | – | – | Linux |
| LKRG, 2018 | – | + | – | + | – | – | Linux |
| LKMG, 2018 | – | + | + | + | + | + | Linux |
| EPTI, 2018 | + | – | + | – | + | – | Linux |
| AllMemPro, 2018 | – | – | + | + | + | + | Windows |
| **MemoryRanger, 2018[8]** | + | + | + | + | + | + | Windows |

[1] PatchGuard protects only limited fields of OS internal structures.
[2] Multics is the first concept of memory isolation for the General Electric (GE) 645 mainframe computer.
[3] HyperCrypt prevents physical attacks such as cold boot via kernel encryption.
[4] Hypernel monitors limited kernel objects.
[5] PrivGuard prevents only privilege escalation attacks.
[6] LAKEED protects memory data related to the keystrokes.
[7] Digtool prevents accessing beyond the bounds of allocated heaps and referencing to freed memory.
[8] MemoryRanger is the proposed memory protection system

# 3. THE MEMORYRANGER: HOW TO RUN DRIVERS IN ISOLATED KERNEL SPACES

The proposed prototype Memory Ranger is based on Intel Virtualization Technology (Intel VT-x) and extended page-table mechanism (EPT). This chapter includes four subsections. The first one deals with how to apply VT-x and EPT to isolate drivers. The second one presents the architecture of MemoryRanger and its implementation. The next section deals with MemoryRanger benchmark results. The final section discusses its limitations.

## 3.1. EPT: The Idea of Memory Isolation

Extended Page Tables (EPT) is an Intel virtualization technology for the memory management unit (MMU) which is designed to virtualize guest physical memory (Intel, 2018).

**EPT Intro.** The central part of this mechanism is the EPT paging structures, which are used during memory translation. According to the section 28.2 of the Intel manual (Intel, 2018), guest-physical addresses are translated by traversing a set of EPT paging structures to produce physical addresses that are used to access memory. Without EPT the guest-physical addresses will be treated as physical addresses and used to access host memory. The address of EPT paging structures is stored in the hypervisor control structure (VMCS).

In a nutshell, EPT plays the role of an intermediary or proxy during memory address translation.

The organization of EPT paging structures is similar to the paging structures in the protected mode.

A detailed analysis of applying EPT to monitor code execution and control memory access has been presented in Section 2.2.2. by Korkin & Tanda (2017). The details of using EPT for protection of allocated data are given in Section 3.1. by Korkin (2018-a).

EPT provides an opportunity to trap and process each access on the memory page by manipulating the content of EPT page-table entry.

It is possible to intercept read, write, and execute memory access attempts by changing the corresponding access attributes on EPT entry as well as redirecting necessary memory access from the original physical page to the fake one by changing page frame number (PFN) value in this EPT entry.

**Memory Isolation using EPT.** The previous research projects show that it is possible to initially allocate fixed EPT paging structures with different access attributes and prevent memory access from kernel-mode drivers by switching between them, for example three EPTs are used in LAKEED by Tian et.al. (2017).

The key idea of MemoryRanger is to dynamically allocate EPT paging structures and update access attributes on EPT page-table entries in real time.

For example, let us consider the following initial scenario. OS Windows is running: OS kernel code and other drivers are loaded into memory. OS kernel code accesses OS structures and other drivers access their memory.

After that, driver A is loaded, allocates the memory data A by calling nt!ExAllocatePoolWithTag routine and accesses this newly allocated memory buffer. Next, Driver B is loaded, allocates data B and accesses this data in a similar way.

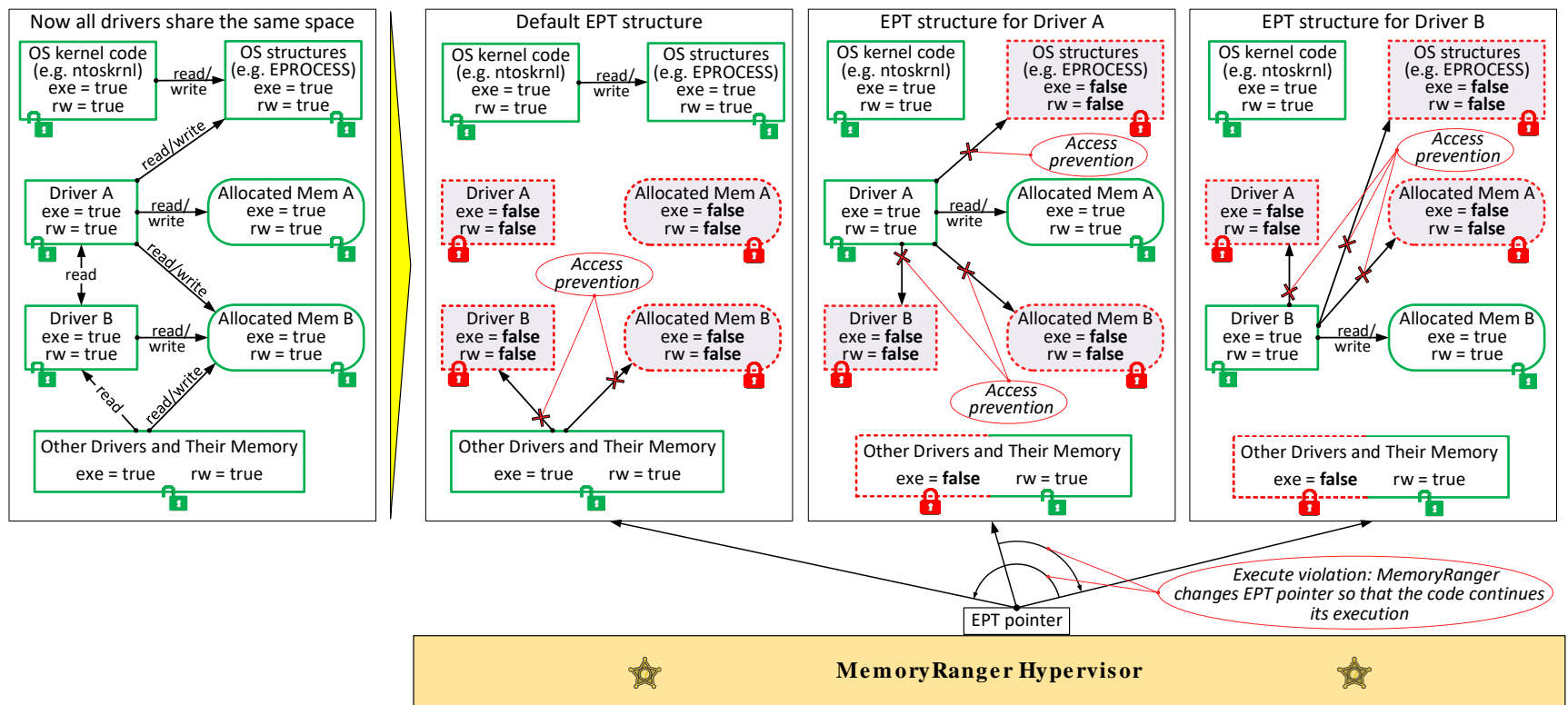

Figure 2. Organization of EPT paging structures to isolate memory of two kernel-mode drivers and their allocated memory

OS kernel code accesses the memory of drivers A and B during its loading. Next, OS code accesses the allocated memory of these drivers during the ExAllocatePoolWithTag call. In this scenario, OS kernel has not been restricted, but it can still be restricted. All the aforementioned three access from drivers to their data are legal and they are marked as horizontal lines, see left part of the Figure 2.

Let us consider the following illegal memory access attempts:

- Driver A does the following:
    o Patches the OS structures;
    o Steals and modifies data B;
    o Dumps the Driver B;
- Driver B does the following:
    o Patches the OS structures;
    o Steals and modifies data A;
    o Dumps the Driver A;
- Other drivers do the following:
    o Steal and modify data A and data B;
    o Dump drivers A and B.

Without MemoryRanger all these illegal memory access attempts are processed without any security reaction. To prevent these attacks MemoryRanger allocates EPT structure for each driver in the following way, see the central and the right parts of the Figure 2.

**Step 1. The Default EPT.** Let us assume that MemoryRanger is loaded as a common driver before Driver A and Driver B will be loaded.

After its loading, MemoryRanger allocates the first EPT structure called the Default EPT. MemoryRanger places OS inside this EPT by setting the memory access attributes: the OS kernel code, OS structures, other drivers and their memory is executable and readable/writable. By default, memory pages of all newly allocated EPT structures are non-executable, but readable and writable.

MemoryRanger receives notifications about drivers loading, the process creation, and memory allocation/deallocation by the third-party driver.

**Step 2. Creating EPT for Driver A.** After trapping the loading of the Driver A, MemoryRanger creates a new EPT structure for Driver A with the following access attributes:

- Memory of Driver A is marked as executable and readable/writable;
- OS kernel code is marked as executable and readable/writable;
- OS structures memory is marked as non-executable and non-readable/non-writable;
- Other drivers and their memory are marked as non-executable, but readable/writable.

Additionally, MemoryRanger updates access attributes for the Default EPT structure:

- Memory of driver A is marked as non-executable and non-readable/non-writable.

**Step 3. Updating two EPTs.** After Driver A allocates memory A, MemoryRanger updates access attributes for two EPTs.

The EPT structure for Driver A gets the following updates:

- Allocated memory A is marked as executable and readable/writable.

The Default EPT structure updates in this way:

- Allocated memory A is marked as non-executable and non-readable/non-writable.

As a result, Driver A is executed and accesses its allocated memory A only in the EPT for Driver A. Access to these memory regions from the Default EPT is forbidden.

**Step 4. Creating EPT for Driver B.** MemoryRanger traps the loading of the Driver B. MemoryRanger creates a new EPT structure for Driver B with the following access attributes:

- Memory of Driver B is marked as executable and readable/writable;
- OS kernel code is marked as executable and readable/writable;
- Memory of Driver A is marked as non-executable and non-readable/non-writable;
- OS structures memory is marked as non-executable and non-readable/non-writable;
- Other drivers and their memory are marked as non-executable, but readable/writable.

Additionally, MemoryRanger updates access attributes for the Default EPT structure and EPT for Driver A in the following way:

- Memory of Driver B is marked as non-executable and non-readable/non-writable.

**Step 5. Updating three EPTs.** After Driver B allocates memory B, MemoryRanger updates all EPTs in the following way.

For the EPT structure for Driver B:

- Allocated memory B is marked as executable and readable/writable.

For the Default EPT structure:

- Allocated memory B is marked as non-executable and non-readable/non-writable.

For the EPT structure for Driver A:

- Allocated memory B is marked as non-executable and non-readable/non-writable.

As a result, Driver B is executed and accesses its allocated memory B only in the EPT for Driver B. Access to these memory regions from the Default EPT and EPT for Driver A is forbidden.

**Step 6. A new process**. MemoryRanger is also notified when a new process is created. After that it reveals the memory address of EPROCESS structure and updates this memory for all EPTs in the following way.

For the Default EPT structure:

- EPROCESS structure is marked as executable and readable/writable.

For the EPT structures for Driver A and Driver B:

- EPROCESS structure is marked as non-executable and non-readable/non-writable.

As a result, only OS kernel and other drivers can access the newly loaded EPROCESS structure from the Default EPT. Access to this memory from all other EPTs and from Driver A (Driver B) is forbidden.

**Step 7. Switching between EPTs.** Windows OS kernel controls drivers' execution using the thread scheduling mechanism. The system's thread scheduler interrupts kernel-mode thread and moves control to another thread (Microsoft, 2004).

Initially, EPT pointer includes the address of the Default EPT. Each time after the OS scheduler moves control to Driver A (or to Driver B) it tries to execute the driver's code and causes an execute EPT violation. MemoryRanger traps this EPT violation, because the corresponding code fragments are marked as non-executable. After trapping, MemoryRanger checks which driver is executed and changes the EPT pointer to the EPT for Driver A (or to the EPT for Driver B) so that the code continues its execution, see Figure 3.

If during execution of Driver A inside the EPT for Driver A the OS scheduler moves control to one of other drivers, its execution leads to the execute EPT violation, because other drivers code fragments are marked as non-executable in EPT for Driver A (and in EPT for Driver B as well). MemoryRanger traps the EPT violation and after deciding which code tries to execute, changes EPT pointer to the Default EPT structure.

In a similar way, MemoryRanger changes EPT pointer if OS kernel code accesses OS structures inside EPT for Driver A (or EPT for Driver B). Memory regions with OS structures are marked as non-readable/non-writeable inside these EPTs and access to the memory always cause EPT violations. This access is granted only inside the Default EPT.

**The Final Step. Preventing illegal access.** Apart from executing EPT violations, MemoryRanger also traps read and write EPT violations. MemoryRanger provides the principle of the minimal privilege: the read and write access to the data is granted only to the drivers, which allocated this data before. The examples of legal access are the following:

In the Default EPT:

- OS kernel core accesses OS structure;
- Other drivers access their memory;

In the EPT for Driver A (in the EPT for Driver B):

- Driver A accesses allocated memory A (Driver B accesses allocated memory B).

All other memory access attempts are trapped and assumed as illegal with one exception. After trapping the EPT violations MemoryRanger decides to grant or prevent an access and implements this decision, see Figure 3. The decision is made according to the following parameters:

- the current value of EPT pointer;
- source address (which code tries to access);
- destination or target address (which data is accessed);
- type of access (read or write).

For illegal access, for example, Driver A tries to access memory of Driver B or Driver B tries to patch OS internal structures, MemoryRanger processes the following steps to prevent an access:

- Redirects access by changing EPT PFN value from the original page to the fake one;
- Allows access to this page by changing EPT memory access attributes;
- Sets Monitor Trap Flag (MTF).

As a result, after a driver reads the fake data the control goes to the hypervisor again. Now MemoryRanger puts the original settings back:

- Restores access by setting EPT PFN value to the original page;
- Blocks access to this page by changing EPT memory access attributes;
- Clears MTF.

These manipulations prevent illegal access to the sensitive data and code.

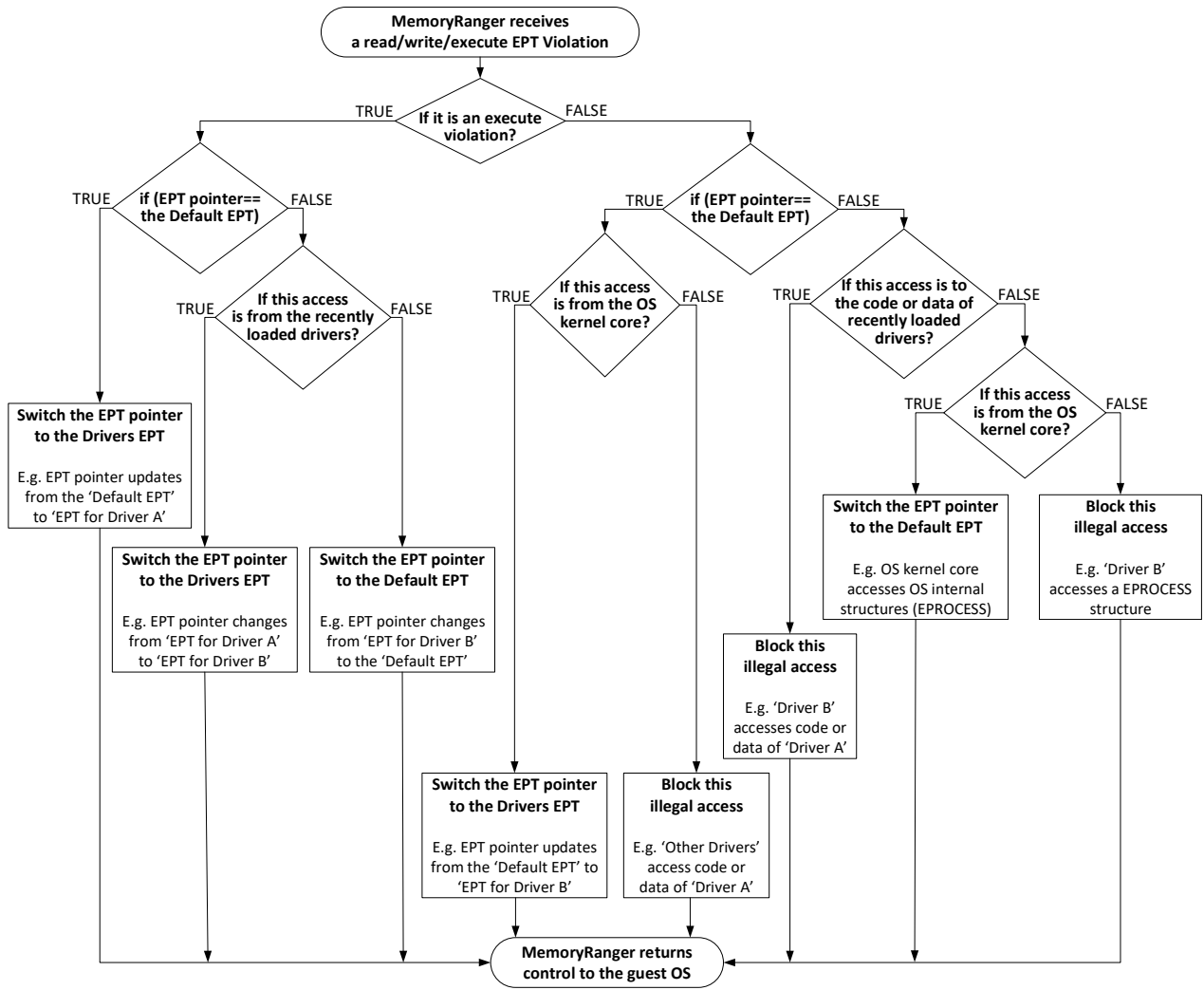

Figure 3. The proposed algorithm of dispatching EPT violations in MemoryRanger (general view)

**The access exception**. The legal read/write memory access can result in EPT violation as well. This exception to the rule is based on memory paging. Windows memory management system can allocate memory for two various drivers in the same 4 kilobyte memory page. As a result, MemoryRanger blocks any access to this memory in all EPT structures.

After the driver tries to access such a memory data, which is allocated by this driver earlier, MemoryRanger traps it. During processing of this violation, MemoryRanger decides that this is legal access and temporarily makes this data accessible:

- Allows access to this page by changing EPT memory access attributes;
- Sets MTF.

After a driver accesses this memory, the control goes to the hypervisor again, and now it implements the following steps to protect memory:

- Blocks access to this page by changing EPT memory access attributes;
- Clears MTF.

These steps help to grant authorized access as well as protecting data buffers, which were allocated at the same memory page.

At the same time, this temporary access granting is very time-consuming. There are several ways of avoiding this issue; one of them is to allocate only page-aligned memory.

**Conclusion.** To sum up, MemoryRanger (MR) isolates drivers execution by leveraging EPT in the following way:

- Initially MR allocates the Default EPT structure. All loaded drivers and OS kernel are executed inside it.
- After a new driver is loaded, MR allocates a new EPT structure with a specific configuration. MR updates all EPT settings so that only this new driver and OS kernel can be executed here.
- Each time the driver allocates memory MR updates all EPT structures again.
- MR updates all EPT structures after a new process has been launched.
- MR provides drives execution by switching between EPT structures.
- MR prevents illegal access attempts by redirecting them to the fake data and restoring EPT settings after each access.
- MR skips legal access to the memory.
- MR isolates code and allocated memory of third-party drivers, which are loaded after it.

Some important details of the implementation details of MemoryRanger are presented in next Section.

### 3.2. Architecture and Implementation of MemoryRanger

MemoryRanger is a bare-metal hypervisor, which is based on hardware virtualization technology VT-x and Extended Page Table (EPT) feature. MemoryRanger hypervisor is loaded using the console application, which starts its legacy driver.

To implement the algorithm from Section 3.1. MemoryRanger needs to process the following:

- starting new processes;
- loading drivers;
- memory allocation and deallocation;
- read/write and execute EPT violations.

To process these events MemoryRanger includes the following parts, see Figure 4:

- Kernel-mode driver with callbacks to be notified about new drivers and processes;
- DdiMon hypervisor, which hooks kernel-mode memory management routines;
- MemoryMonRWX hypervisor, which handles EPT violations and EPT structures;
- Memory Access Policy (MAP), which is a kind of brain for processing all the events.

For each of the following notifications: drivers loading, launching processes, memory allocation and deallocation, MemoryRanger adds the corresponding data structures to the lists, and updates EPT structures.

The first component is a kernel-mode driver, which registers two driver-supplied callback routines using PsSetCreateProcessNotifyRoutineEx and PsSetLoadImageNotifyRoutine to receive notifications about processes creation (MSDN, 2018-a) and drivers loading (MSDN, 2018-b).

Whenever a process is created the corresponding callback routine creates the structure EPROCESS_PID and sends it to the MAP. This structure includes two fields:

- process ID;
- vector of addresses and sizes of EPROCESS memory regions, which are needed to be protected.

MAP adds this structure to the list and updates EPT structures using MemoryMonRWX hypervisor.

In a similar way, another callback routine receives notifications about drivers loading. After a new driver is loaded, this callback creates the ISOLATED_MEM_ENCLAVE structure and sends it to the MAP. Here is this structure:

- address of newly allocated EPT paging structure for this driver;
- driver's image base address;
- driver's image end address, which is a sum of base address and image size;
- vector of allocated memory pools.

MAP adds the ISOLATED_MEM_ENCLAVE to the corresponding list, creates a new EPT paging structure and updates all other EPT structures.

The second component processes kernel APIs. MemoryRanger considers that third-party drivers allocate memory using ExAllocatePoolWithTag routine and free using ExFreePoolWithTag. MemoryRanger intercepts these kernel API calls using DdiMon. It is a hypervisor-based project, which leverages EPT facilities to install stealth hooks by Tanda (2016).

DdiMon receives a notification about memory allocation and sends this data to the MAP:

- address of code, which allocates the buffer;
- address and size of allocated memory pool.

MAP receives it and creates the ALLOCATED_POOL structure. Next MAP finds ISOLATED_MEM_ENCLAVE structure corresponding to the driver, which allocates this memory, and adds ALLOCATED_POOL into the vector 'drv_allocs' from this structure. Finally, MAP updates EPT structures to take into account a newly allocated memory buffer.

In a similar way, DdiMon processes memory deallocation and removes ALLOCATED_POOL structure. This scheme helps to supply up-to-date information about which memory pools have been allocated by which driver.

The third component is MemoryMonRWX (Korkin & Tanda, 2017), which controls access to the memory in real time. This hypervisor-based component handles read, write, and execute violations and sends the following data about each EPT violation to the MAP:

- the current value of EPT pointer;
- source address (which code tries to access);
- destination address (which data is accessed);
- type of access (read, write, or execute).

The final component is MAP, which receives this data and makes the decision using the lists of EPROCESS_PID, ISOLATED_MEM_ENCLAVE, ALLOCATED_POOL. MAP will then grant, or will prevent memory access, or will change EPT pointer according to the algorithm, see Figure 3.

MemoryRanger is developed using Microsoft Visual C++ 2015 with integrated Windows Driver Kit (WDK). It is tested using Vmware Workstation 14 and Windows 10 1709 64-bit. The source code of MemoryRanger is found here Korkin (2018-b).

**Demos.** The proposed MemoryRanger architecture implements all steps to isolate drivers' execution and it has been successfully tested in two scenarios. In the first demo, MemoryRanger protects both code and allocated data of third-party drivers from illegal access. In the second scenario, MemoryRanger prevents privilege escalation attack (Korkin, 2018-c).

The next section will cover the benchmark assessment results of MemoryRanger.

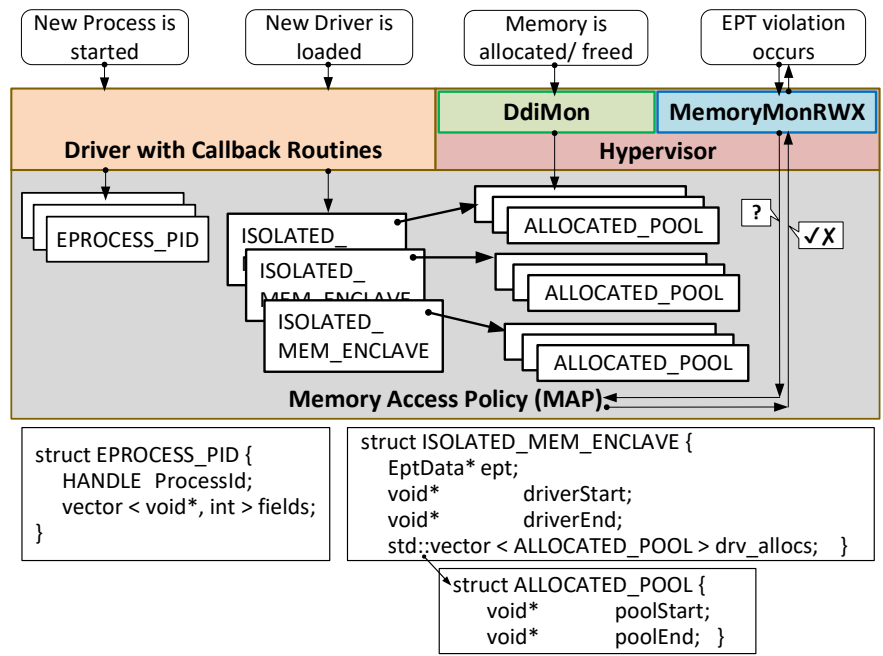

Figure 4 MemoryRanger Architecture with four parts: the driver, DdiMon, MemoryMonRWX, and MAP

## 3.3. Benchmark Results

This section covers the benchmark results of MemoryRanger and compares them with the AllMemPro, which is the nearest competitor.

The benchmark was evaluated in the following way. A driver allocates memory buffer and measures the duration of memory access to the data using Time Stamp Counter (TSC).

The benchmark was calculated in four cases:

- without hypervisor with enabled memory cache;
- without hypervisor and disabled cache;
- with AllMemPro hypervisor;
- with MemoryRanger hypervisor.

The calculated values of average memory access latency and its deviation are presented in Table 2. All the details related to the number of measurements and calculations are here Korkin (2018-a).

The first three cases are similar with my previous research see Korkin (2018-a).

The nearest competitor of MemoryRanger is AllMemPro (Korkin, 2018-a). It uses only one EPT structure to protect allocated data and traps both types of access attempts: legal and illegal ones. As a result, each memory access attempt causes significant time degradation.

MemoryRanger excludes this drawback by using separate EPT structures for each driver. This helps to trap only illegal access and skip the legal ones, whose latency values are measured during benchmark.

MemoryRanger is about three times faster than the nearest competitor, and it is slower by half than the OS without hypervisor with disabled cache.

This time degradation happens for two reasons:

- During the time measurement the OS schedule is switching EPT pointer between the EPT for Driver X and the Default EPT.
- The changing EPT pointer results in TLB flushing and further memory access requires page-walk, which is time consuming.

The first one is based on the fact that Windows is a preemptive multitasking OS and cannot be avoided.

The second issue with the TLB flushing can be partially mitigated. The authors of EPTI hypervisor show that it is possible to speed up hypervisor by avoiding TLB flush after changing EPT pointer. This idea will be checked in further research.

I can conclude that MemoryRanger has good benchmark results and these results can be improved.

## 3.4. Discussion and Limitations

MemoryRanger is a proof of concept prototype and has several limitations.

MemoryRanger has similar limitations with AllMemPro developed by Korkin (2018-a):

- Indirect memory access;
- Self-protection;
- Protection of memory with shared access;
- Page file mechanism and forcing page-out;
- Direct access to the physical memory;
- Joint work with Windows 10 UEFI version;
- SGX technology and Virtual Secure Mode.

One of them is the protection of data buffers, which have to be accessed from both user-mode and kernel-mode code, for example IRP and MDL data. MemoryRanger does not protect this data.

Table 2 Time evaluation

| No. | Cases | Memory Access Latency, TSC ticks |
|---|---|---|
| 1 | without hypervisor with enabled cache | 70±2 |
| 2 | without hypervisor with disabled cache | 100.000±4.000 |
| 3 | with AllMemPro by Korkin (2018-a) | 500.000±10.000 |
| 4 | with MemoryRanger | 170.000±7.000 |

## 4. CONCLUSIONS

To sum up I would like to highlight the following:

1. Windows OS kernel is vulnerable to malware attacks and security researchers are trying to fill this gap.
2. MemoryRanger protects kernel-mode code and allocated data from illegal access by executing drivers in separate enclaves.
3. MemoryRanger provides confidentiality and integrity for the memory of third-party drivers and the OS internal structures.
4. MemoryRanger achieves a low performance overhead due to allocating a kernel-mode enclave for each driver.
5. MemoryRanger is a hypervisor-based solution with flexible architecture, which does not require the drivers' source code.

## 5. FUTURE STEPS

MemoryRanger is a very promising project and here are my five steps for its future development.

### 5.1. Spectre and Meltdown Attacks

MemoryRanger seems to prevent data leakage from kernel-mode via side channel attacks based on hardware vulnerabilities such as Meltdown attacks presented by Lipp et.al. (2018).

To prevent Meltdown attack, MemoryRanger can isolate user-mode and kernel-mode spaces by allocating additional EPT paging structures for user space, like EPTI (Hua et.el., 2018), and encrypt memory pages with sensitive data, like HyperCrypt by Götzfried et.al. (2016).

MemoryRanger can isolate user-mode data from being stolen or modified by kernel-mode malware.

### 5.2. Restriction of OS Kernel to Prevent Exploitation

There are several hypervisor-based projects, which do not restrict OS kernel, for example LKMG. As a result, after malware exploits OS kernel core vulnerability, it can access sensitive data in memory.

MemoryRanger can restrict the OS kernel core. The current version of MemoryRanger allocates a separate EPT paging structures for each driver, which includes the OS kernel code and the corresponding driver's code. The OS kernel code can be restricted by excluding it from the driver's EPT, after the driver has been loaded.

### 5.3. Next Areas for Drivers Isolation: File System, Registry, Network, Devices

Windows security does not prevent illegal access from kernel-mode drivers to the file system, registry, and network. As a result, a malware driver can read/write/modify, create/delete files, registry and network data, which are processed by user-mode applications or other drivers. Also, a malware driver can access camera, microphone, and other devices in an unauthorized way.

MemoryRanger can implement the corresponding access rules to isolate file system, registry, network, and devices from being accessed by malware drivers.

### 5.4. Creating Access Rules from Drivers' Source Code During Compilation

The current version of MemoryRanger does not provide security shared access to the memory. As a result, only the driver, which allocates the data, can access it. One of the possible ways to tackle this issue is to generate memory access rules using the drivers' source code.

MemoryRanger can use memory access policies, which are generated during the drivers' compilation. Usually, drivers, which allocate data for shared access, are compiled in the same environment, for example, one Visual Studio solution includes two projects with such drivers.

### 5.5. Integrate into Windows Kernel

Finally, by integrating MemoryRanger into Windows OS kernel, we can significantly improve data protection from both software and hardware attacks. It is time to take a step forward and protect data of more than one billion Windows users all around the world.

## 6. ACKNOWLEDGMENTS

I would like to thank Michael Chaney, ERAU alumni, Cyber-Research Editor, USA, for reviewing this research and providing constructive feedback, which significantly contributed to improving the quality of this paper.

I am also very grateful to Ivan Nesterov (i.nesterow@gmail.com), head of the R&D

laboratory, Russia, for his invaluable contribution, ideas, and support. His main research areas include information security, high-performance computing, distributed storage systems, database design, and high availability applications. His software design experience includes applications on hybrid CPU/GPU special-purpose architectures for telecommunication and cryptography, distributed visualization and data science.

# 7. REFERENCES

[1] ACSC (2018, May). Hardening Microsoft Windows 10 version 1709 Workstations. Australian Cyber Security Centre (ACSC). Retrieved from https://acsc.gov.au/publications/protect/Hardening_Win10.pdf

[2] Azab, A.M., Ning, P., Shah, J., Chen, Q., Bhutkar, R., Ganesh, G., Ma, J., & Shen, W. (2014). Hypervision Across Worlds: Real-time Kernel Protection from the ARM TrustZone Secure World. In Proceedings of the 2014 ACM SIGSAC Conference on Computer and Communications Security (CCS '14). ACM, New York, NY, USA, pp. 90-102. DOI: https://doi.org/10.1145/2660267.2660350

[3] Brookes, S., Denz, R., Osterloh, M., & Thayer, S.T. (2016, April). ExOShim: Preventing Memory Disclosure using Execute-Only Kernel Code. In Proceedings of the 11th International Conference on Cyber Warfare and Security, ICCWS'16, pp. 56-66. Retrieved from http://thayer.dartmouth.edu/tr/reports/tr15-001.pdf

[4] Chen, Q., Azab, A.M., Ganesh, G., & Ning., P. (2017). PrivWatcher: Non-bypassable Monitoring and Protection of Process Credentials from Memory Corruption Attacks. In Proceedings of the 2017 ACM on Asia Conference on Computer and Communications Security (ASIA CCS '17). ACM, New York, NY, USA, pp. 167-178. DOI: https://doi.org/10.1145/3052973.3053029

[5] Corbató, F. J., & Vyssotsky, V. A. (1965). Introduction and Overview of the Multics System. In Proceedings of the AFIPS Fall Joint Computer Conference, FJCC. Volume 27, Part 1. ACM, New York, NY, USA, pp. 185-196. DOI: http://dx.doi.org/10.1145/1463891.1463912

[6] Desimone, J., & Landau, G. (2018, August 9). Kernel Mode Threats and Practical Defenses. In Proceedings of the BlackHat USA Conference. Las Vegas, Nevada. Retrieved from https://i.blackhat.com/us-18/Thu-August-9/us-18-Desimone-Kernel-Mode-Threats-and-Practical-Defenses.pdf

[7] Feng, X., Yang, Q., Shi L., & Wang, Q. (2018). BehaviorKI: Behavior Pattern Based Runtime Integrity Checking for Operating System Kernel, 2018 IEEE International Conference on Software Quality, Reliability and Security (QRS), Lisbon, Portugal, pp. 13-24. DOI: https://doi.org/10.1109/QRS.2018.00015

[8] Gadaleta F., Nikiforakis N., Mühlberg J.T., & Joosen W. (2012). HyperForce: Hypervisor-enForced Execution of Security-Critical Code. In Proceedings of the 27th IFIP TC 11 International Information Security Conference. DOI: https://doi.org/10.1007/978-3-642-30436-1_11

[9] Ge, X., Vijayakumar, H., & Jaeger, T. (2014). SPROBES: Enforcing Kernel Code Integrity on the TrustZone Architecture. In Proceedings of the 3rd Workshop on Mobile Security Technologies (MoST). San Jose, CA. Retrieved from https://arxiv.org/abs/1410.7747

[10] Ghosh, S. (2017, May 15). Windows XP is still the third most popular operating system in the world. Business Insider. Retrieved from http://uk.businessinsider.com/windows-xp-third-most-popular-operating-system-in-the-world-2017-5

[11] Götzfried, J., Dörr, N., Palutke, R., & Müller, T. (2016). HyperCrypt: Hypervisor-Based Encryption of Kernel and User Space. In the Proceedings of the 11th International Conference on Availability, Reliability and Security (ARES). Salzburg, Austria. DOI: https://doi.org/10.1109/ARES.2016.13

[12] Han, S., Kang, J. Shin, W., Kim, H., & Park, E. (2017). Myth and Truth about Hypervisor-Based Kernel Protector: The Reason Why You Need Shadow-Box. Retrieved from https://www.blackhat.com/docs/asia-17/materials/asia-17-Han-Myth-And-Truth-about-Hypervisor-Based-Kernel-Protector-

The-Reason-Why-You-Need-Shadowbox-wp.pdf

[13] He, Y., Zheng, X., Zhu, Z., & Shi, G. (2017). TZ-SSAP: Security-Sensitive Application Protection on Hardware-Assisted Isolated Environment. Security and Privacy in Communication Networks. Springer International Publishing. pp. 538-556. DOI: https://doi.org/10.1007/978-3-319-59608-2_30

[14] Hofmann, O., Dunn, A., Kim, S., Roy, I., & Witchel, E. (2011). Ensuring operating system kernel integrity with OSck. In Proceedings of the 16th International Conference on Architectural Support for Programming Languages and Operating Systems (ASPLOS XVI). ACM, New York, NY, USA, 279-290. DOI: https://doi.org/10.1145/1950365.1950398

[15] Hofmann, O., Kim, S., Dunn, A., Lee M., & Witchel, E. (2013, March 16–20). InkTag: Secure Applications on an Untrusted Operating System. In Proceedings of the 18th International Conference on Architectural Support for Programming Languages and Operating Systems (ASPLOS '13). ACM, New York, NY, USA, 265-278. DOI: https://doi.org/10.1145/2451116.2451146

[16] Hua, Z., Du, D., Xia Y., Chen H., & Zang, B. (2018). EPTI: Efficient Defence against Meltdown Attack for Unpatched VMs. In Proceedings of the USENIX Annual Technical Conference (ATC). Boston, MA. pp. 255-266. Retrieved from https://www.usenix.org/conference/atc18/presentation/hua

[17] Intel. (2018, May). Intel® 64 and IA-32 Architectures Software Developer's Manual. Combined Volumes: 1, 2A, 2B, 2C, 2D, 3A, 3B, 3C, 3D and 4. Order Number: 325462-067US. Retrieved from https://software.intel.com/en-us/download/intel-64-and-ia-32-architectures-sdm-combined-volumes-1-2a-2b-2c-2d-3a-3b-3c-3d-and-4

[18] Korkin, I. (2018-a). Hypervisor-Based Active Data Protection for Integrity and Confidentiality of Dynamically Allocated Memory in Windows Kernel. Paper presented at the Proceedings of the 13th Annual Conference on Digital Forensics, Security and Law (CDFSL), University of Texas at San Antonio (UTSA), San Antonio, Texas, USA. pp. 7-38 Retrieved from https://igorkorkin.blogspot.com/2018/03/hypervisor-based-active-data-protection.html

[19] Korkin, I. (2018-b). MemoryRanger source code. GitHub repository. Retrieved from https://github.com/IgorKorkin/MemoryRanger

[20] Korkin, I. (2018-c). MemoryRanger Demos. The Attack and the Attack Prevention. [Video file]. Retrieved from https://www.youtube.com/watch?list=PL0Aerbf3kwULpVhoHyjMUeUFLwnvur5iu&v=HNxc-tjy3QA

[21] Korkin, I., & Tanda, S. (2017, May 15-16). Detect Kernel-Mode Rootkits via Real Time Logging & Controlling Memory Access. Paper presented at the Proceedings of the 12th Annual Conference on Digital Forensics, Security and Law (CDFSL), Embry-Riddle Aeronautical University, Daytona Beach, Florida, USA. pp. 39-78. Retrieved from https://igorkorkin.blogspot.com/2017/03/memorymonrwx-detect-kernel-mode.html/

[22] Kwon, D., Oh, K., Park, J., Yang, S., Cho, Y., Kang, B., & Paek, Y. (2018, June). Hypernel: a Hardware-Assisted Framework for Kernel Protection Without Nested Paging. In Proceedings of the 55th Annual Design Automation Conference (DAC '18). ACM, New York, NY, USA, Article 34, 6 pages. DOI: https://doi.org/10.1145/3195970.3196061

[23] Lin, T., Chan, E. M., Farivar, R., Mallick, N., Carlyle, J. C., David, F. M., & Campbell, R. H. (2007). iKernel: Isolating Buggy and Malicious Device Drivers Using Hardware Virtualization Support. In Proceedings - DASC 2007: Third IEEE International Symposium on Dependable, Autonomic and Secure Computing. pp. 134-142. DOI: https://doi.org/10.1109/ISDASC.2007.4351398

[24] Lipp, M., Schwarz, M, Gruss, D., Prescher, T., Haas, W., Fogh, A., Horn, J., Mangard, S., Kocher, P., Genkin, D., Yarom, Y., & Hamburg, M. (2018). Meltdown: Reading Kernel Memory from User Space. 27th USENIX Security Symposium. Retrieved from https://meltdownattack.com/meltdown.pdf

[25] Liu, P., Tian, D., & Xiong, X. (2011). Practical Protection of Kernel Integrity for Commodity

OS from Untrusted Extensions. In Proceedings of the 18th Annual Network and Distributed System Security Symposium (NDSS). Retrieved from http://citeseerx.ist.psu.edu/viewdoc/download?doi=10.1.1.477.1924&rep=rep1&type=pdf

[26] Manes, V., Jang, D., Ryu, C., & Kang, B. (2018, May). Domain Isolated Kernel: A Lightweight Sandbox for Untrusted Kernel Extensions. Elsevier Computers & Security. Volume 74. pp. 130–143. DOI: https://doi.org/10.1016/j.cose.2018.01.009

[27] Microsoft. (2004). Scheduling, Thread Context, and IRQL. Windows Hardware and Driver Central. Microsoft Download Center. Retrieved from http://download.microsoft.com/download/e/b/a/eba1050f-a31d-436b-9281-92cdfeae4b45/irql_thread.doc

[28] MSDN (2018-a). PsSetCreateProcessNotifyRoutineEx routine. Kernel-Mode Driver Reference. Retrieved from https://docs.microsoft.com/en-us/windows-hardware/drivers/ddi/content/ntddk/nf-ntddk-pssetcreateprocessnotifyroutineex

[29] MSDN (2018-b). PsSetLoadImageNotifyRoutine routine. Kernel-Mode Driver Reference. Retrieved from https://docs.microsoft.com/en-us/windows-hardware/drivers/ddi/content/ntddk/nf-ntddk-pssetloadimagenotifyroutine

[30] Oh, M. (2018, September 27). Return of the Kernel Rootkit Malware (on Windows 10). In Proceedings of the BlueHat v18 Security Conference. Redmond, Washington, USA. Retrieved from https://www.slideshare.net/MSbluehat/bhv18-return-of-the-kernel-rootkit-malware-on-windows-10

[31] Pan, J., Yan, G., & Fan, X. (2017). Digtool: A Virtualization-Based Framework for Detecting Kernel Vulnerabilities. In Proceedings of the 26th USENIX Conference on Security Symposium (SEC'17). pp. 149-165. USENIX Association, Berkeley, CA, USA. Retrieved from https://www.usenix.org/system/files/conference/usenixsecurity17/sec17-pan.pdf

[32] Pomonis, M., Petsios, T., Keromytis, A.D., Polychronakis, M., & Kemerlis, V.P. (2017). kR^X: Comprehensive Kernel Protection against Just-In-Time Code Reuse. In Proceedings of the 12th European Conference on Computer Systems (EuroSys '17). ACM, New York, NY, USA, 420-436. DOI: https://doi.org/10.1145/3064176.3064216

[33] Qiang, W., Yang, J., Jin H., & Shi, X. (2018, August 21). PrivGuard: Protecting Sensitive Kernel Data from Privilege Escalation Attacks. In Proceedings of the IEEE Access, vol. 6, pp. 46584-46594. DOI: https://doi.org/10.1109/ACCESS.2018.2866498

[34] Shirole, S. (2014, April 15). Performance Optimizations for Isolated Driver Domains. Virginia Polytechnic Institute and State University. Master of Science in Computer Science & Applications. Blacksburg, Virginia. Retrieved from https://vtechworks.lib.vt.edu/bitstream/handle/10919/49107/Shirole_SM_T_2014.pdf

[35] Srivastava, A., Erete, I., & Giffin, J. (2009). Kernel Data Integrity Protection via Memory Access Control. Tech. Rep. GT-CS-09-04, Georgia Institute of Technology. Retrieved from https://smartech.gatech.edu/handle/1853/30785

[36] Tanda, S. (2016). Monitor Device Driver Interfaces (DDIMon). GitHub repository. Retrieved from https://github.com/tandasat/DdiMon

[37] Tian, D., Jia, X., Chen, J., & Hu. C. (2017). An Online Approach for Kernel-level Keylogger Detection and Defense. Journal of Information Science and Engineering. Volume 2. Number 2. pp. 445-461. Retrieved from http://jise.iis.sinica.edu.tw/JISESearch/pages/View/PaperView.jsf?keyId=155_2013

[38] Tian, D., Xiong, X., Hu, C., & Liu, P. (2011). Policy-Centric Protection of OS Kernel from Vulnerable Loadable Kernel Modules. In Information Security Practice and Experience - 7th International Conference, ISPEC 2011, pp. 317-332. Springer-Verlag, Berlin, Heidelberg. DOI: https://doi.org/10.1007/978-3-642-21031-0_24

[39] Tian, D., Xiong, X., Hu, C., & Liu, P. (2018, March 8). A Policy-Centric Approach to Protecting OS Kernel from Vulnerable LKMs.

Journal of Software: Practice and Experience. DOI: https://doi.org/10.1002/spe.2576

[40] Wang, J.P., Zhao, P., & Ma, H.T. (2017, July). HACS: A Hypervisor-Based Access Control Strategy to Protect Security-Critical Kernel Data. 2$^{nd}$ International Conference on Computer Science and Technology (CST 2017). Guilin, China, DOI: https://doi.org/10.12783/dtcse/cst2017/12516

[41] Wang, X., Chen Y., Wang, Z., Qi Y., & Zhou, Y. (2015). SecPod: a Framework for Virtualization-based Security Systems. USENIX Annual Technical Conference (ATC'15). USENIX Association. Retrieved from https://www.usenix.org/conference/atc15/technical-session/presentation/wang-xiaoguang

[42] Warren, T. (2018, March 30). Microsoft is ready for a world beyond Windows. The Verge. Retrieved from https://www.theverge.com/2018/3/30/17179328/microsoft-windows-reorganization-future-2018

[43] Yi, H., Cho, Y., Paek, Y., & Ko. K., (2017, September 1). DADE: A Fast Data Anomaly Detection Engine for Kernel Integrity Monitoring. The Journal of Supercomputing. Springer US. DOI: https://doi.org/10.1007/s11227-017-2131-6

[44] Yosifovich, P., Ionescu, A., Russinovich, M., & Solomon, D. (2017). Windows Internals, Part 1, 7$^{th}$ ed. Redmond, Washington: Microsoft Press.

[45] Zabrocki, A. (2018). LKRG – Linux Kernel Runtime Guard ver 0.4. Retrieved from https://www.openwall.com/lkrg/

[46] Zhang, F., Wang, J., Sun, K., & Stavrou, A. (2014). HyperCheck: A Hardware-Assisted Integrity Monitor. IEEE Transactions on Dependable and Secure Computing, 11(4). https://doi.org/10.1109/TDSC.2013.53